\documentclass{ws-procs975x65}

\begin{document}

\title{Microlensing towards LMC and M31}

\author{Philippe Jetzer and Sebastiano Calchi Novati\footnote{\uppercase{W}ork 
supported by the \uppercase{S}wiss 
\uppercase{N}ational \uppercase{S}cience \uppercase{F}oundation
and by the \uppercase{T}omalla \uppercase{F}oundation.}}

\address{
Institute of Theoretical Physics\\
           University of Z\"urich, 
Winterthurerstarsse 190, 8057 Z\"urich, Switzerland\\
E-mails: jetzer@physik.unizh.ch, novati@physik.unizh.ch}

\maketitle

\abstracts{
The nature and the location of the lenses discovered in the
microlensing surveys done so far towards the LMC remain unclear.
Motivated by these questions we computed the optical depth 
for the different intervening populations and
the number of expected events for self-lensing, using a recently drawn 
coherent picture of the geometrical structure and dynamics of the
LMC. By comparing the
theoretical quantities with the values of the observed events it
is possible to put some constraints on the location and the nature
of the MACHOs. 
Clearly, given the large uncertainties and the few
events at disposal it is not yet possible to draw sharp conclusions,
nevertheless we find that up to 
3-4 MACHO events might be due to lenses in LMC, which are most probably
low mass stars, but that hardly all events can be due to
self-lensing. The most
plausible solution is that the events observed so far are due to
lenses belonging to different intervening populations: low mass
stars in the LMC, in the thick disk, in the spheroid and some true
MACHOs in the halo of the Milky Way and the LMC itself.
We report also on
recent results of the SLOTT-AGAPE and POINT-AGAPE collaborations
on a search for microlensing events in direction of the Andromeda galaxy,
by using the pixel method.  The detection of 4 microlensing events, 
some likely to be due to self--lensing, is discussed. One microlensing light curve is shown to be 
compatible with a binary lens. The present analysis still
does not allow us to draw conclusions on the MACHO content of the
M31 galaxy.}

\section{Introduction}

Since Paczy\'nski's original proposal\cite{pacz}
gravitational microlensing has been proben to be
a powerful tool for the detection of the dark matter
component in galactic haloes in the form of MACHOs.
Searches in our Galaxy towards LMC\cite{alcock00a,eros} show
that up to 20\% of the halo could be formed
by objects of around $M \sim 0.4\,M_\odot$.

However, the location and the nature of the microlensing events found so
far towards the Large Magellanic Cloud (LMC) is still a matter of
controversy. The MACHO collaboration found 13 to 17 events in 5.7
years of observations, with a mass for the lenses estimated to be
in the range $0.15 - 0.9 ~M_{\odot}$  assuming a
standard spherical Galactic halo\cite{alcock00a} and
derived an optical depth of $\tau= 1.2^{+0.4}_{-0.3} \times
10^{-7}$. The EROS2
collaboration\cite{milsztajn} announced the discovery
of 4 events based on three years of observation but
monitoring about twice as much stars as the MACHO collaboration.
The MACHO collaboration monitored primarily 15 deg$^2$ in the
central part of the LMC, whereas the EROS2 experiment covers a
larger solid angle of 64 deg$^2$ but in less crowded fields. The
EROS2 microlensing rate should thus be less affected by
self-lensing. This might be the reason for the fewer events seen
by EROS2 as compared to the MACHO experiment.

The hypothesis for a self-lensing component
was discussed by several authors 
\cite{sahu,salati,evans98,zhao}.
The analysis of Jetzer et al. \cite{jetzer02} and 
Mancini et al. \cite{mancini} 
has shown that probably the observed
events are distributed among different galactic components (disk,
spheroid, galactic halo, LMC halo and self-lensing). This means that
the lenses do not belong all to the same population and their
astrophysical features can differ deeply one another.

Some of the events found by the MACHO team are most probably due
to self-lensing: the event MACHO-LMC-9 is a double lens with
caustic crossing\cite{alcock00b} and its proper motion is
very low, thus favouring an interpretation as a double lens within
the LMC. The source star for the event MACHO-LMC-14 is double\cite{alcock01b} 
and this has allowed to conclude that the
lens is most probably in the LMC. The expected LMC self-lensing
optical depth due to these two events has been estimated to lie
within the range\cite{alcock01b} $1.1-1.8\times10^{-8}$,
which is still below the expected optical depth for self-lensing
even when considering models giving low values for the optical
depth.
The event LMC-5 is due to a disk lens\cite{alcock01c} and
indeed the lens has even been observed with the HST. 
The other stars which have been
microlensed were also observed but no lens could be detected, thus
implying that the lens cannot be a disk star but has to be either
a true halo object or a faint star or brown dwarf in the LMC
itself.

Thus up to now the question of the location of the observed MACHO events
is unsolved and still subject to discussion. Clearly, with much more
events at disposal one might solve this problem
by looking for instance at their spatial distribution. 
To this end a correct knowledge of the structure
and dynamics of the luminous part of the LMC is essential, and 
we take advantage of the new picture drawn by van der Marel et al. 
\cite{marel01a,marel01b,marel02}. 

Searches towards M31, nearby and similar to our Galaxy,
have also been proposed\cite{crotts92,agape93,jetzer94}.
This allows to probe a different line of sight in our Galaxy,
to globally test M31 halo and, furthermore,
the high inclination of the M31 disk is expected
to provide a strong signature (spatial distribution) for halo microlensing signals.

Along a different direction, results of a microlensing survey towards M87, 
where one can probe both the M87 and the Virgo cluster haloes, 
have also been presented\cite{m87}.

For extragalactic targets, due to the distance, the sources for microlensing signals
are not resolved. This claims for an original technique, the \emph{pixel method}, 
the detection of flux variations of unresolved sources\cite{agape97,agape99,tom96},
the main point being that one follows flux variations of every pixel
in the image instead of single stars. 

We discuss here the results from 
two different survey of M31 aimed at the detection
of microlensing events, carried out by the  SLOTT-AGAPE\cite{mdm1,mdm2}
and by the POINT-AGAPE collaborations\cite{point01,point03}.
The WeCapp\cite{wecapp,wecapp03} 
and the MEGA\cite{mega} collaborations have also
presented a handful of microlensing events.

\section{LMC model}
\label{morphology}

In a series of three 
interesting papers \cite{marel01a,marel01b,marel02}, a
new coherent picture of the geometrical structure and dynamics of
LMC has been given. In the following 
we adopt this model and use 
the same coordinate systems and notations as in van der
Marel. 
We consider an elliptical isothermal flared disk
tipped by an angle $i = 34.7^{\circ} \pm 6.2^{\circ}$ as to the
sky plane, with the closest part in the north-east side. The
center of the disk coincides with the center of the bar 
and its distance from us is $D_{0} = 50.1 \pm
2.5 \, \mathrm{kpc}$. We take a bar mass
$M_{\mathrm{bar}}=1/5\,M_{\mathrm{disk}}$ with
$M_{\mathrm{bar}}+M_{\mathrm{disk}}=M_{\mathrm{vis}}=2.7 \times
10^{9}\, \mathrm{M}_{\odot}$.

The vertical distribution of stars in an isothermal disk is
described by the $\mathrm{sech}^{2}$ function; therefore the
spatial density  of the disk is modeled by:
\begin{equation}
\rho_{\mathrm{d}}=\frac{N\, M_{\mathrm{d}}}{4 \pi\, q\,
R_{\mathrm{d}}^{2}\, \zeta_{\mathrm{d}}(0)}\;
\mathrm{sech}^2\left( \frac{\zeta}{\zeta_{\mathrm{d}}(R)}\right)
 e^{-{\frac{1}{R_{\mathrm{d}}}}\sqrt{
\left(\frac{\xi}{q}\right)^2+
{\eta}^2}}~,
\end{equation}
where $q = 0.688$ is the ellipticity factor,
$R_{\mathrm{d}}=1.54\,\mathrm{kpc}$ is the scale length of the
exponential disk, $R$ is the radial distance from the center on
the disk plane. ${\zeta_{\mathrm{d}}(R)}$ is the \textit{flaring}
scale height, which rises from 0.27 kpc to 1.5 kpc at a distance
of 5.5 kpc from the center\cite{marel02}, and is
given by
$${\zeta_{\mathrm{d}}(R)}=0.27+1.40
\,\tanh \left(\frac{R}{4}\right)~.
$$
$N = 0.2765$ is a normalization factor that takes into
account the flaring scale height.

In a first approach\cite{jetzer02} 
we have described the bar by a Gaussian density profile
following Gyuk et al. \cite{gyuk}, whereas in a following
paper\cite{mancini} we choose, instead, a
bar spatial density  that takes into account its boxy shape\cite{zhao}:

\begin{equation}
\rho_{\mathrm{b}}=\frac{2\,M_{\mathrm{b}}}{\pi^{2} \,
R_{\mathrm{b}}^{2}\,\, \Xi_{\mathrm{b}}}\, e^{
-\left(\frac{\Xi}{\, \Xi_{\mathrm{b}}}\right)^{2}}
\,e^{-\,\frac{1}{R_{\mathrm{b}}^{4}}
\left(\Upsilon^{2}+\,\zeta^{2}\right)^{2}},
\end{equation}
where $\Xi_{\mathrm{b}}=1.2\,\mathrm{kpc}$ is the scale length of
the bar axis,  $R_{\mathrm{b}}=0.44\,\mathrm{kpc}$ is the scale
height along a circular section (for a more detailed discussion
and definition of the coordinate system see \cite{mancini}).

\begin{figure}
 \resizebox{\hsize}{!}{\includegraphics{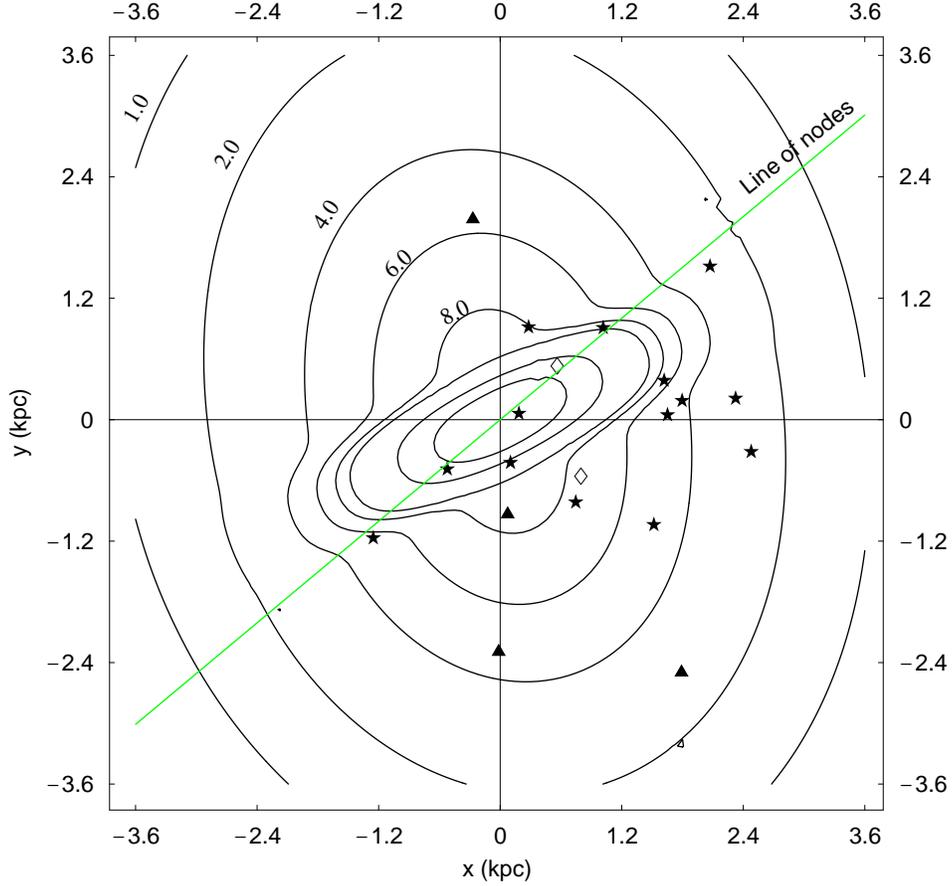}}
 \caption{Projection on the sky plane ($x-y$ plane) of the column density
 of the LMC disk and bar. The numerical values on the contours are in $10^{7}
 \mathrm{M}_{\odot}\mathrm{kpc}^{-2}$ units.
 The three innermost contours correspond to $10$, $20$ and $30\times 10^{7}
 \mathrm{M}_{\odot}\mathrm{kpc}^{-2}$. The locations of the MACHO
 (black stars and empty diamonds) and EROS
 (triangles) microlensing candidates are also shown.}
 \label{column-density-plot}
\end{figure}

The column density, projected on the $x-y$ sky plane is plotted in
Fig. \ref{column-density-plot}, giving a global view of the LMC
shape for a terrestrial observer, together with the positions of
the microlensing events detected by the MACHO (filled 
stars and empty diamonds) and EROS (filled triangles)
collaborations, and the direction of the line of nodes. The
maximum value of the column density, $41.5\times 10^{7}~
\mathrm{M}_{\odot}\,\mathrm{kpc}^{-2}$, is assumed in the center
of LMC. 

We use two different models to describe the halo profile density:
a spherical halo and an ellipsoidal halo. The
values of the parameters have been chosen so that the models have
roughly the same mass within the same radius. In the spherical model
we neglect the tidal effects due to our
Galaxy, and we adopt  a classical pseudo-isothermal spherical  density
profile:
\begin{equation}
\rho_\mathrm{h,S}=\rho_{\mathrm{0,S}}
\left(1+\frac{R^{2}}{a^{2}}\right)^{-1} \theta(R_{\mathrm{t}}-R),
\end{equation}
where $a$ is the LMC halo core radius,  $\rho_{0,\mathrm{S}}$ the
central density, $R_{\mathrm{t}}$ a cutoff radius and $\theta$ the
Heaviside step function. We use $a=2$ kpc.
We fix the value for the mass of the halo within a radius of 8.9
kpc  equal to\cite{marel02} $5.5\times10^{9}~\mathrm{M}_{\odot}$ 
that implies $\rho_{0,\mathrm{S}}$ equal to
$1.76\times10^{7}\,~\mathrm{M}_{\odot}\,\mathrm{kpc}^{-3}$. Assuming
a halo truncation radius\cite{marel02}, $R_{\mathrm{t}}= 15$ kpc, 
the total mass of the halo is $\approx
1.08\times10^{10}\,~\mathrm{M}_{\odot}$.

For the galactic halo we assume a spherical model with density
profile given by:
\begin{equation}
\rho_\mathrm{GH}=\rho_{\mathrm{0}}
\frac{R_{\mathrm{C}}^{2} + R_{\mathrm{S}}^{2}}{R_{\mathrm{C}}^{2}+R^{2}},
\end{equation}
where $R$ is the distance from the galactic center,
$R_{\mathrm{C}} = 5.6 \,\,\mathrm{kpc}$ is
the core radius, $R_{\mathrm{S}} = 8.5 \, \,\mathrm{kpc}$ is the
distance of the Sun from the galactic center and
$\rho_{\mathrm{0}} = 7.9\times 10^{6}\,
~\mathrm{M}_{\odot}\,\mathrm{kpc}^{-3}$ is the mass density in the
solar neighbourhood.

\section{LMC optical depth} \label{tau}
%
The computation is made by weighting the optical depth with
respect to the  distribution of the source stars along the line of
sight (see Eq.(7) in Jetzer et al. \cite{jetzer02}):
\begin{equation}\label{weightedOD}
\tau = {\frac{4\pi G}{c^{2}}}
{\frac{\int_{0}^{\infty}\left[\int_{0}^{D_{\mathrm{os}}}
{\frac{D_{\mathrm{ol}}(D_{\mathrm{os}}-D_{\mathrm{ol}})}
{D_{\mathrm{os}}}}\rho_{\mathrm{l}}
\,dD_{\mathrm{ol}}\right]\rho_{\mathrm{s}}\,
dD_{\mathrm{os}}}{\int_{0}^{\infty}\rho_{\mathrm{s}}\,
dD_{\mathrm{os}}}}\;.
\end{equation}
$\rho_{\mathrm{l}}$ denotes the mass density of the lenses,
$\rho_{\mathrm{s}}$ the mass density of the sources,
$D_{\mathrm{ol}}$ and $D_{\mathrm{os}}$, respectively, the distance
observer-lens and observer-source.

\begin{figure}
 \resizebox{\hsize}{!}{\includegraphics{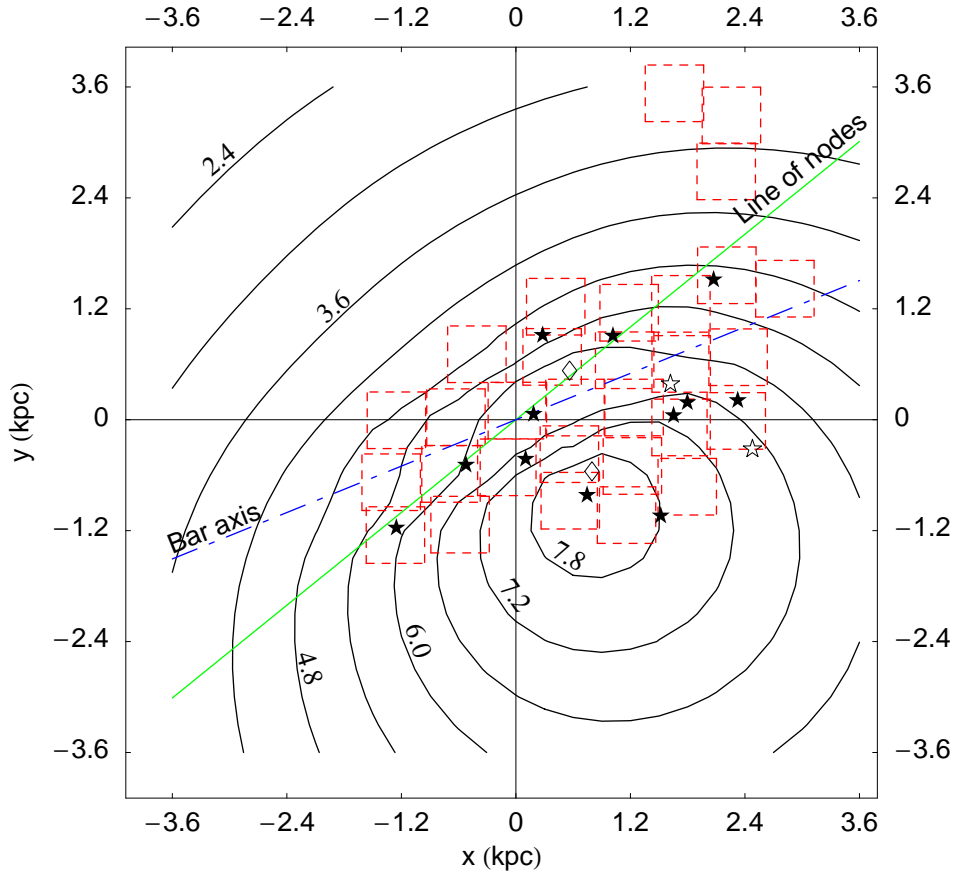}}
 \caption{Spherical halo model: contour map of the optical depth for
 lenses in the LMC halo. The locations of the
 MACHO fields and of the microlensing candidates are also shown.
 The numerical values are in $10^{-8}$ units.}
 \label{LMCHS}
\end{figure}

In Fig. \ref{LMCHS} we report the optical depth
contour maps for lenses belonging to the halo of LMC
in the case of spherical model
in the hypothesis that all the LMC dark halo consists of compact
lenses. The ellipsoidal model leads to similar results\cite{mancini}.
A striking feature of the map is the strong near-far
asymmetry.

For the spherical model, the maximum value of the optical depth,
$\tau_{\mathrm{max,S}} \simeq 8.05 \times 10^{-8}$, is assumed in
a point falling in the field number 13, belonging to the fourth
quadrant, at a distance of $\simeq 1.27$ kpc from the center. The
value in the point symmetrical with respect to the center,
belonging to the second quadrant and falling about at the upward
left corner of the field 82, is $\tau_{\mathrm{S}} \simeq 4.30
\times 10^{-8}$. The increment of the optical depth is of the
order of $\approx 87\%$,  moving from the nearer to the farther
fields. 

\begin{figure}
\resizebox{\hsize}{!}{\includegraphics{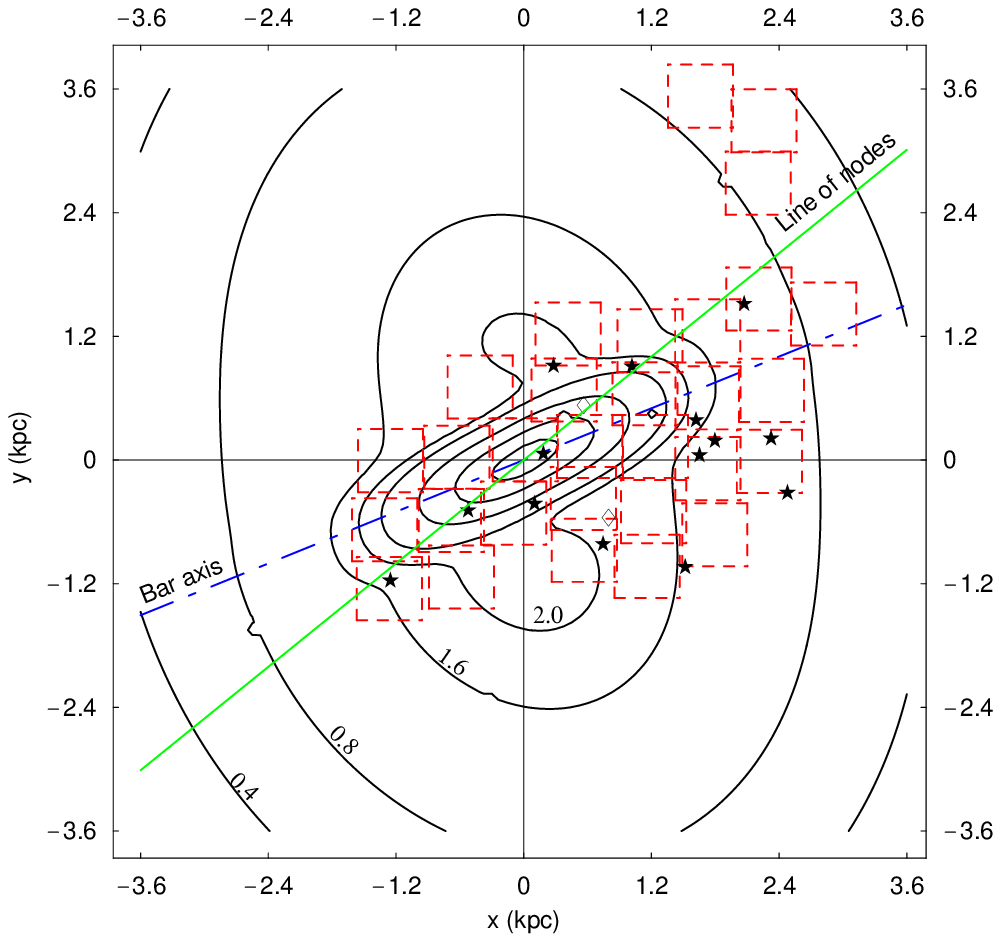}}
\caption{Contour map of the optical depth for self-lensing.  The
 locations of the MACHO fields and of the microlensing candidates
 are also shown. The numerical values are in $10^{-8}$ units.
 The innermost contours correspond to
 values $2.4\times 10^{-8}$, $3.2\times 10^{-8}$, $4.0\times 10^{-8}$
 and $4.6\times 10^{-8}$ respectively.}
\label{SL}
\end{figure}

In Fig. \ref{SL} we report the optical depth contour map for
self-lensing, i. e. for events where both the sources and the
lenses belong to the disk and/or to the bulge of LMC. As expected,
there is almost no near-far asymmetry
and the maximum value of the optical depth, $\tau_{\mathrm{max}}
\simeq 4.80 \times 10^{-8}$, is reached in the center of LMC. The
optical depth then  rapidly decreases, when moving, for instance,
along a line going through the center and perpendicular to the
minor axis of the elliptical disk, that coincides also with the
major axis of the bar. In a range of about only $0.80
\,\mathrm{kpc}$ the optical depth  quickly falls to $\tau \simeq
2 \times 10^{-8}$, and afterwards it decreases slowly to lower
values.
%

\section{LMC self-lensing event rate}
An important quantity, useful for the physical interpretation of
microlensing events, is the distribution
$\frac{\mathrm{d}\Gamma}{\mathrm{d}\,T_{\mathrm{E}}}$, the differential
rate of microlensing events with respect to the Einstein time
$T_{\mathrm{E}}$. In particular it allows us to estimate the
expected typical duration and their expected number. 
We evaluated the microlensing rate in the self-lensing
configuration, i. e. lenses and sources both in the disk and/or in
the bar of LMC. We have taken into account the transverse motion of
the Sun and of the source stars.
We assumed that, to an observer comoving with the LMC center, the
velocity distribution of the source stars and lenses have a
Maxwellian profile, with spherical symmetry.

In the picture of van der Marel et al. within a distance of
about 3 kpc from the center of LMC, the velocity dispersion
(evaluated for carbon stars) along the line of sight can be
considered  constant, $\sigma_{\mathrm{los}}=20.2\pm .5$ km/s.
Most of the fields of the MACHO collaboration fall within this
radius and, furthermore, self-lensing events are in any case
expected to happen in this inner part of LMC. Therefore, 
we adopted this value, even if we are aware that the
velocity dispersion of different stellar populations in the LMC
varies in a wide range, according to the age of the stellar
population: $\simeq 6$ km/s for the youngest population, until
$\simeq 30$ km/s for the older ones\cite{gyuk}.

We need now to specify the form of the number density. Assuming
that the mass distribution of the lenses is independent of their
position\cite{derujula} 
in LMC ({\it factorization hypothesis}), the lens number
density per unit mass is given by
\begin{equation}
{\frac{\mathrm{d}n_{\mathrm{l}}}{\mathrm{d}
\mu}}={\frac{\rho_{\mathrm{d}}+\rho_{\mathrm{b}}}{\mathrm{M}_{\odot}}}\,
{\frac{\mathrm{d} n_{0}}{\mathrm{d} \mu}},
\end{equation}
where we use ${\frac{\mathrm{d} n_{0}}{\mathrm{d} \mu}}$  as given
in Chabrier\cite{chabrier} ($\mu = M/M_{\odot}$). 
We consider both the power law and the
exponential initial mass functions\footnote{We have used the same
normalization as in Jetzer et al. \cite{jetzer02}
with the mass varying in the range
0.08 to 10 M$_{\odot}$.}. However, we find that our results do not
depend strongly on that choice and hereafter, we will discuss the
results we obtain by using the exponential IMF only.

Let us note that, considering the experimental conditions for the
observations of the MACHO team, we use as  range 
for the lens masses  $0.08 \le\mu \le 1.5$. The lower limit
is fixed by the fact that the lens must be a star in LMC, while
the upper limit is fixed by the requirement that the lenses are
not resolved stars\footnote{We have checked that the results are
insensitive to the precise upper limit value.}.

We compute the ``field exposure'', 
$E_{\mathrm{field}}$, defined, as in Alcock et al. \cite{alcock00a}, 
as the product of the number of distinct light curves per
field by the relevant time span, paying attention to eliminate
the field overlaps; moreover we calculate the distribution
${\frac{\mathrm{d}\Gamma}{\mathrm{d}T_{\mathrm{E}}}}$ along the line
of sight pointing towards the center of each field. In this way we
obtain the number of  expected events for self-lensing, field
by field, given by
\begin{equation}
N_{\mathrm{SL,field}}=E_{\mathrm{field}}\int_{0}^{\infty}\,
{\frac{\mathrm{d}\Gamma}{\mathrm{d}T_{\mathrm{E}}}}\, 
E(T_{\mathrm{E}})\, \mathrm{d}\,T_{\mathrm{E}} \; ,
\end{equation}
where $E(T_{\mathrm{E}})$ is the detection efficiency.

Summing over all fields we find that the expected total number of self-lensing events is 
$\sim 1.2$, while we would get $\sim 1.3$ with the the double power law IMF; in both cases
altogether 1-2 events\cite{mancini}. Clearly, taking also into account the
uncertainties in the parameter used following the van der Marel
model for the LMC the actual number could also be somewhat higher
but hardly more than our upper limit 
estimate of about 3-4 events given in\cite{jetzer02}.

\subsection{Self-lensing events
discrimination}

%
It turns out that, in the framework of the LMC
geometrical structure and dynamics outlined above, a
suitable statistical analysis allows us to exclude from the
self-lensing population a large subset of the detected events. To
this purpose, assuming  all the 14 events as self-lensing, we
study the  scatter plots  correlating the self-lensing expected
values of some meaningful microlensing variables with the measured
Einstein time or with the self-lensing optical depth. In this way
we can show that a large subset of events is clearly incompatible with
the self-lensing hypothesis.

We have calculated the self-lensing distributions
$\left({\frac{\mathrm{d}\Gamma}{\mathrm{d}T_{\mathrm{E}}}}\right)_{\varepsilon}$
of the rate of microlensing events with respect to the Einstein
time $T_{\mathrm{E}}$, along the lines of sight towards the 14
events found by the MACHO collaboration, in the case of a Chabrier
exponential type IMF. 
With these distributions we have calculated the modal
$T_{\mathrm{E,mod}}$, the median $T_{\mathrm{E},\,50\,\%}$ and the
average $<T_{\mathrm{E}}>$ values of the Einstein time.

\begin{figure}
\resizebox{\hsize}{!} {\includegraphics{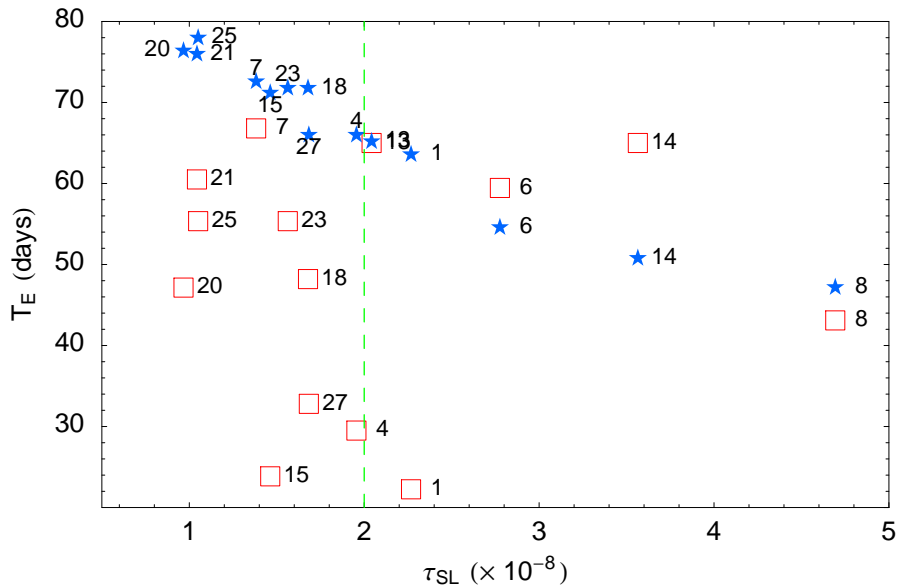}}
{\caption{Scatter plot of the observed (empty boxes) values of the
Einstein time and of the expected values of the median
$T_{\mathrm{E},50\, \%}$ (filled stars), with respect to the
self-lensing optical depth evaluated along the directions of the
events.} \label{tevstau}}
\end{figure}

In Fig. \ref{tevstau} we report on the $y$--axis the observed
values of $T_\mathrm{E}$ (empty boxes) as well as the expected
values for self-lensing of the \emph{median}
$T_{\mathrm{E}\,,50\,\%}$ (filled stars) evaluated
\emph{along the directions of the events}. On the $x$--axis we
report the value of the self-lensing optical depth calculated
towards the event position; the optical depth is  growing going
from the outer regions towards the center of LMC according to the
contour lines shown in Fig. \ref{SL}. An interesting feature
emerging clearly is the \emph{decreasing} trend of the  expected
values of the median $T_{\mathrm{E}\,,50\,\%}$, going from the
outside fields with low values of $\tau_{\mathrm{SL}}$ towards the
central fields with higher values of $\tau_{\mathrm{SL}}$. The
variation of the stellar number density and the flaring of the LMC
disk certainly contributes to explain this behaviour.

We now tentatively identify two subsets of events: the nine
falling outside the contour line $\tau_{\mathrm{SL}} = 2 \times
10^{-8}$ of Fig. \ref{SL} and the five falling inside. In the
framework of van der Marel et al. LMC geometry, this  contour line
includes almost fully the LMC bar and two ear shaped inner regions
of the disk, where we expect self-lensing events to be located with
higher probability.

We note that, at glance, the two clusters have a clear-cut
different collective behaviour: the measured Einstein times of the
first $9$ points fluctuate around a median value of 48 days, very
far from the expected values of the median $T_\mathrm{E}$, ranging
from 66 days to 78 days, with an average value of 72 days. On the
contrary, for the last 5 points, the measured Einstein times
fluctuate around a median value of 59 days, very near to the
average value 56 days of the expected medians, ranging from 47
days to 65 days. Let us note, also, the somewhat peculiar position
of the event LMC--1, with a very low value of the observed
$T_{\mathrm{E}}$; 
most probably this event is homogeneous to the set at left of the
vertical line in Fig. \ref{tevstau} and it has to be included in
that cluster.

This plot gives a first clear evidence that, in the framework of
van der Marel et al. LMC geometry, the self-lensing events have
to be searched among the cluster of events with
$\tau_{\mathrm{SL}}\,>\, 2\times 10^{-8}$, and at the same time
that the cluster of the $9$ events including LMC--1  belongs, very
probably, to a different population.

Moreover, when looking at the spatial distribution of the events
one sees a clear near-far asymmetry in the van der Marel
geometry; they are concentrated along the extension of the bar and
in the south-west side of LMC. Indeed, we have performed a statistical 
analysis of the spatial distribution of the events, which clearly 
shows that the observed asymmetry is greater than the one expected
on the basis of the observational strategy\cite{mancini}.

\section{Pixel lensing towards M31 with MDM data}

The SLOTT-AGAPE collaboration has been using data collected on the 1.3m 
McGraw-Hill Telescope at the MDM observatory, Kitt Peak (USA). Two
fields, $17'\times 17'$ wide each, on the opposite side (and including) the bulge are observed
(centered in $\alpha=$ 00h 43m 24s, $\delta  = 41^{\!\circ}12'10''$
(J2000) ``Target'', on the far side of M31, and $\alpha=$ 00h 42m 14s,
$\delta  =41^{\!\circ}24'20''$ (J2000) ``Control''). Two filters, 
similar to standard $R$ and $I$ Cousins, have been used in order to test achromaticity. 
Furthermore, this particular colour information gives 
the chance of having a better check on red
variable stars, which can contaminate the search
for microlensing events. Observations have been carried out in a two years
campaign, from October 1998 to the end of December 1999.
Around 40 (20) nights of observations are available
in the Target and Control field respectively. 

To cope with photometric and seeing variations
we follow the ``superpixel photometry''\cite{agape97,mdm1} approach,
where one statistically calibrate the flux of each image
with respect of a chosen reference image.
In particular, the seeing correction is based
on an empirical linear correction of the flux,
and we do not need to evaluate the PSF of the image.

The search for microlensing events is carried out in two steps.
Through a statistical analysis on the light curve
significant flux variations above the baseline are detected,
then we perform a shape analysis on the selected light curve,
$\sim 10^3$, to distinguish between microlensing and other
variable stars. 

The background of variable sources is a main problem
for pixel lensing searches of microlensing signals. 
First, the class of stars to which we are in principle most sensitive are the red giants, 
for which a large fraction are variable stars 
(regular or irregular). Second, as looking for \emph{pixel}
flux variations, it is always possible to collect (in the same pixel)
light from more than one source whose flux is varying.
Thus, in the analysis, one is faced with two problems: 
large-amplitude variable sources whose signal can mimic a microlensing signal,
and variable sources of smaller amplitude whose signal can 
give rise to non-gaussian fluctuations
superimposed on the background or on other physical variations.

In a first analysis\cite{mdm1} we followed a conservative
approach to reduce the impact of these problems. Severe criteria
in the shape analysis with respect to the Paczy\'nski fit were adopted
(with a stringent cut for the $\chi^2$)
and, furthermore, candidates  with both a long timescale
($t_{1/2}>40$ days) and a red colour ($(R-I)_C>1$) were excluded, since these most likely
originate from variable stars.

\begin{figure}[ht]
\centerline{
\epsfxsize=3.65in
\epsfbox{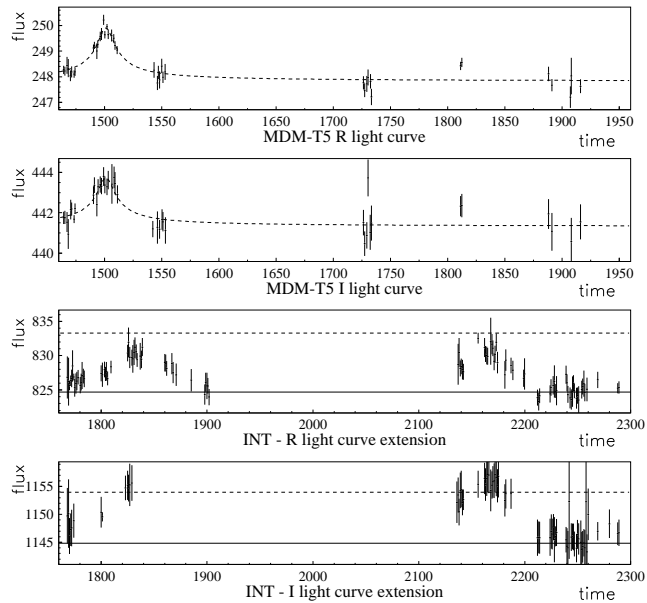}
}
\caption{MDM T5 light curve together with its extension into the INT data.
On the $y$ axis, flux is in ADU/s; on the $x$ axis, time is in days, with the origin in J-2449624.5 (both data sets). 
For the MDM light curves the dashed line represent the result of the Paczy\'nski fit.
For the INT light curves, shown together with the solid line representing
the baseline is a dashed line representing the level of the maximum deviation of flux found on the corresponding MDM light curve.}
\label{mdmcl-1}
\end{figure}

In this way 10 variations compatible
with a microlensing (time width in the range 15-70 days,
and flux deviation at maximum all above $\Delta R\sim 21.5$) 
were selected. However, due to the rather poor sampling and the short baseline,
the uniqueness bump requirement could not be proben efficiently.
A successive analysis\cite{mdm2} on the INT extension of these
light curves then shows that all these variations are indeed
due to variable sources and rejected as microlensing candidates. 
Indeed, in the same position, a variation with compatible time width and flux deviation
is always found on INT data. In Fig. \ref{mdmcl-1} we show
one MDM flux variation (T5) from this selection, nicely fitting
a Paczy\'nski light curve, then its extension on the INT data 
where it is clearly seen that the bump does repeat with the same shape,
showing that this is actually a variable source.

\begin{figure}[ht]
\centerline{
\epsfxsize=3.5in
\epsfbox{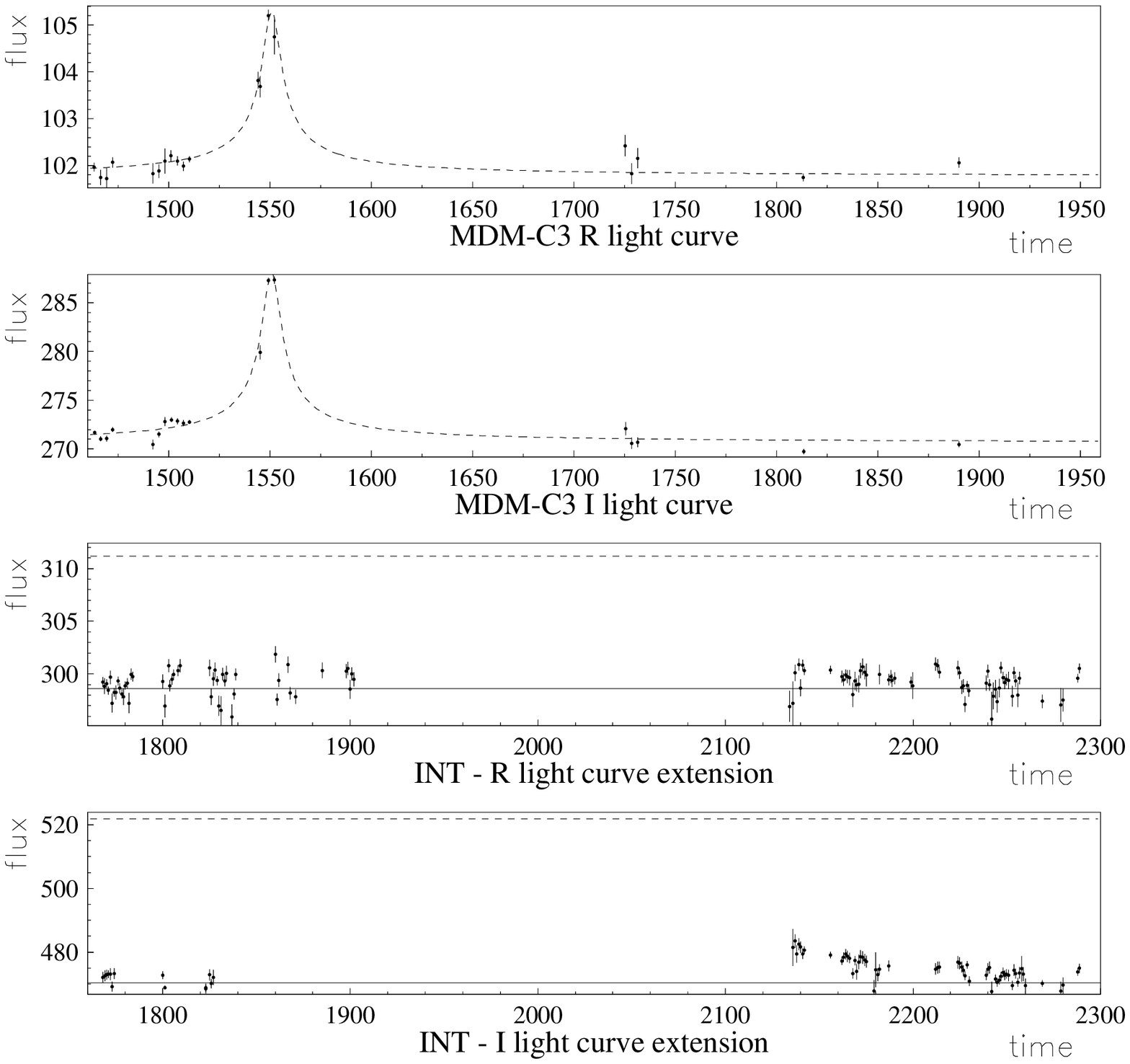}}   
\caption{MDM C3 light curves together with their extension into the INT data. Symbols as in Fig. \ref{mdmcl-1}.}
\label{mdmcl-2}
\end{figure}

A second analysis is then carried out where we relax
the criteria introduced to characterize the shape, as this
has proben not to efficiently reject variable stars and 
indeed could introduce a bias against real microlensing
events whose light curve might be disturbed by some
non gaussian noise, and, on the other hand, we restrict the allowed space of physical
parameters, in particular we consider only relatively short
(time width less than 20 days) flux variations (this range of
parameter space being consistent with what expected 
on the basis on Monte Carlo simulations\cite{mdm1}).

As an outcome, out of further 8 detected flux variations,
INT vetting allows to firmly exclude 5 as microlensing,  
leaving 2 light curves for which this test is considered inconclusive
and 1 lying in a region of space not covered by the INT field
(with $t_{1/2}\in (13,20)$ days and $\Delta R_{max} \in (21.0,21.8)$).

By ``inconclusive'' it is meant that a flux variation
is detected at the same position on INT data,
but where the comparison of the time width and the flux deviation
added to the rather poor sampling along the bump 
do not allow to conclude sharply on the uniqueness test,
leaving open the possibility of the detection
of a microlensing light curve superimposed on (the light curve of)
a variable star (Fig. \ref{mdmcl-2}).

\section{Microlensing events towards M31 with INT data}

The POINT-AGAPE collaboration\cite{point01} is carrying out a survey of M31
by using the Wide Field Camera (WFC) on the 2.5 m INT telescope.
Two fiels, each of $\sim 0.3$ deg$^2$ are observed. The observations
are made in three bands close to Sloan $g',\,r',\,i'$. We
report here on the results from the analysis of 143
nights collected in two years between August 1999 and January 2001.
As described for MDM data, superpixel photometry is performed
to bring all the images to the same reference one,
then a similar analysis for the search of microlensing
candidates is carried out.

\begin{figure}[ht]
\centerline{
\epsfbox{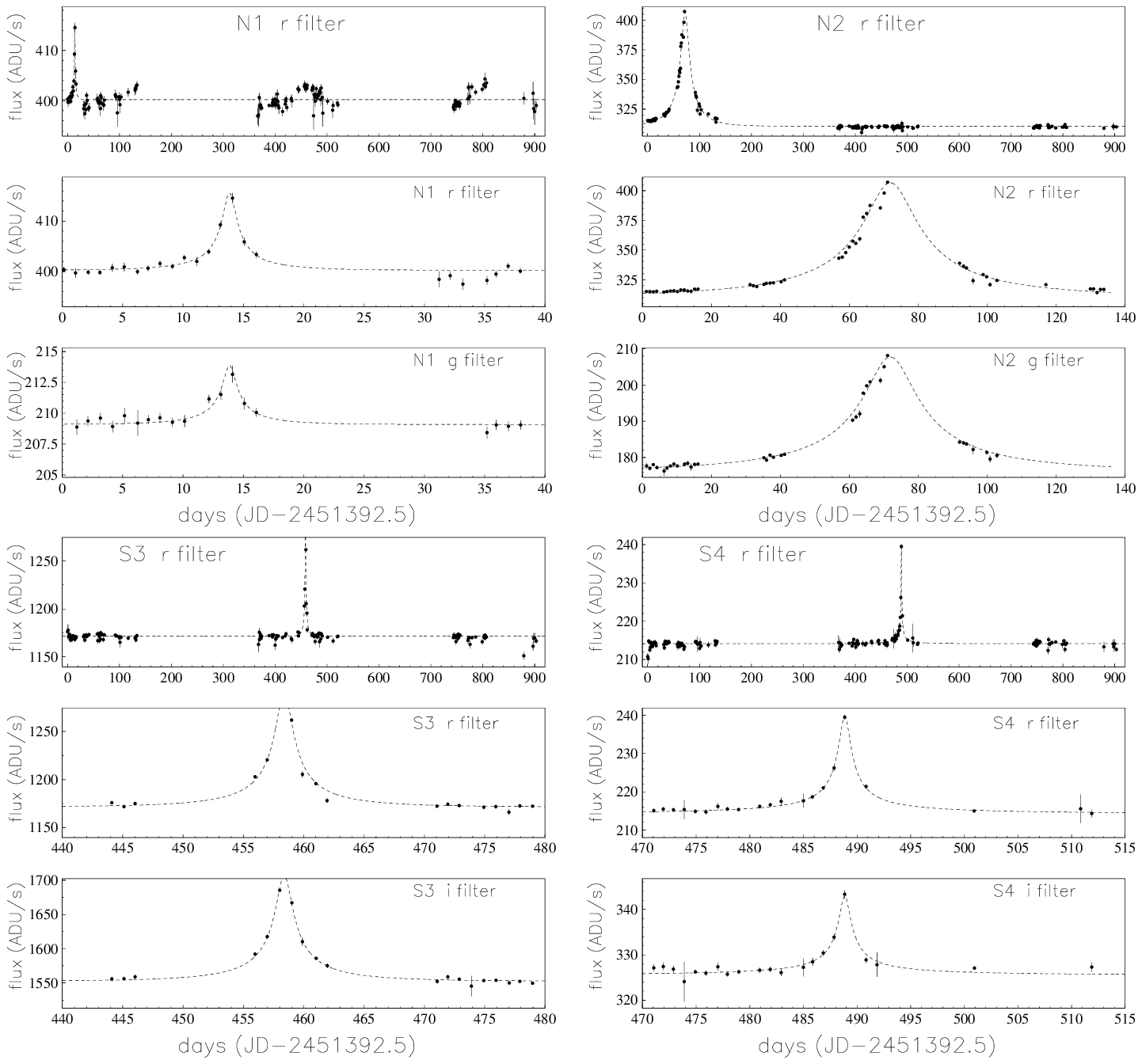}}
\caption{Three years data light curves for the 4 POINT-AGAPE microlensing events.}
\label{int4}
\end{figure}

A first analysis\cite{point03} is made with the aim
to detect short ($t_{1/2} < 25$ days) and bright variations
($\Delta R < 21$ at maximum amplification),
compatible with a Paczy\'nski signal.
The first requirement is suggested
by the results on the predicted characteristics
of microlensing events of a Monte Carlo simulation
of the experiment. As an outcome,
four light curves are detected, whose characteristics
are summarised in Table \ref{tab}, and whose light curve
are shown in Fig. \ref{int4} (with a third year data
added). We stress that their
signal is incompatible with any known variable star,
therefore it is safe to consider these as viable
microlensing events.

\begin{table}[h]
\tbl{Characteristics of the four microlensing events
detected by the POINT-AGAPE collaborations. $d$ is the projected
distance from the center of M31.\label{tab}}
{\begin{tabular}{@{}ccccc@{}} \toprule
 & PA-99-N1 & PA-99-N2 & PA-00-S3 &
PA-00-S4 \\ \colrule
$\alpha$ (J2000) & 00h42m51.4s & 00h44m20.8s & 00h42m30.5s & 00h42m30.0s \\
$\delta$  (J2000) & $41^\circ\, 23'\, 54''$ & $41^\circ\, 28'\, 45''$ &
           $41^\circ\, 13'\, 05''$ & $40^\circ\, 53'\, 47''$\\
$d$      & $7'\, 52''$ & $22'\, 03''$ & $4'\, 00''$ & $22'\, 31''$\\
$t_{1/2}$ (days) & $1.8\pm 0.2$ & $21.8\pm 0.2$ & $2.2\pm 0.1$ & $2.1\pm 0.1$  \\
$\Delta R_{max}$ & $20.8\pm 0.1$ & $19.0\pm 0.2$ & $18.8\pm 0.2$ & $20.7\pm 0.2$  \\
\botrule
\end{tabular}}
\end{table}

\begin{figure}[ht]
\centerline{
\epsfbox{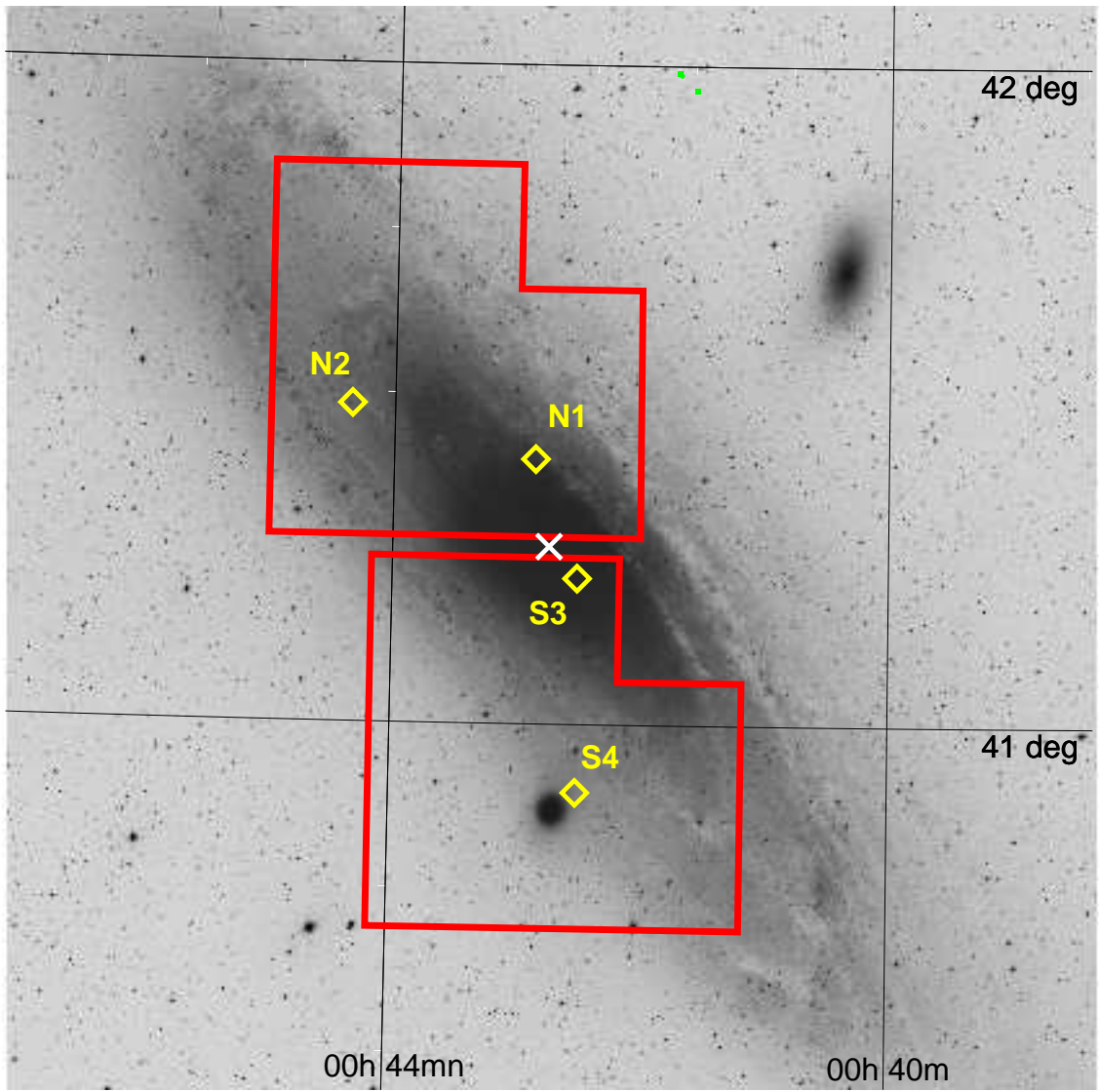}
}
\caption{Positions of the four microlensing events projected on M31. The red lines
show the contour of the observed fields. Note that S4 lies just next to M32.}
\label{int-pos}
\end{figure}

Once a microlensing event is detected it is important,
given the aim to probe the halo content in form
of MACHO, to find out its origin, namely, whether it is due to
self-lensing within M31 or to a MACHO. This is not
straightforward. The spatial distribution of the events
is an important tool, but still unusable given the small statistic.
The observed characteristics of the variations to some extent
can give a hint on the nature of the lens, but again,
the small number of detected events so far makes this 
approach rather unviable. However, we stress that the detection 
of some self-lensing event, as they are expected to be found (their existence
being predicted only on the basis of the rather well known
luminous component of M31), is essential to assess
the efficiency of the analysis. In the following,
starting from their spatial position (Fig. \ref{int-pos}) we briefly
comment on each of the detected events.

PA-99-N1: For this event it has been possible
to identify the source on HST archival images. 
The knowledge of the flux of the unamplified source allows
to break the degeneracy between the Einstein
crossing time and the impact parameter
for which one obtains the values $t_E=9.7 \pm 0.7$
and $u_0 = 0.057 \pm 0.004$. The baseline
shows two secondary bumps: they are due to
a variable star lying some 3 pixels away from the 
microlensing variation. The position of this event,
$7'\,52''$ from the center of M31, makes
unlikely the hypothesis of bulge-bulge self-lensing.
The lens can be either a MACHO (with equal chance
in the M31 or the Milky Way halo) or a low mass
($\sim 0.2\,M_\odot$) disk star with the source
lying in the bulge. The first  case is more likely 
assuming a halo fraction in form of MACHOs above 20\%.

PA-99-N2: This variation lies at some $22''$
away from the center of M31, therefore
is an excellent microlensing MACHO  candidate.
However, this variation turns out to be
almost equally likely to be due to
disk--disk self lensing. This light curve is particularly
interesting because it shows clear deviations
from a Paczy\'nski shape, while remaining
achromatic (and unique) as expected for a microlensing event.
We recall that this shape is characteristic
for variations where the point-like (source and lens)
and uniform motion hypothesis hold. After exploring\cite{point03a}
different explanations, it is found that the observations
are consistent with an unresolved RGB or AGB star in M31
being microlensed by a binary lens, with
a mass ratio of $\sim 1.2\times 10^{-2}$. An analysis
of the relative optical depth shows that a halo
lens (whose mass is estimated to lie in the range
0.009-32 $M_\odot$) is more likely
than a stellar lens (with mass in this case expected
in the range 0.02-3.6 $M_\odot$) provided that 
the halo mass fraction in form of compact objects
is at least around $15\%$.

PA-00-S3: This event is the nearest found so far
from the center of M31 ($d=4'\,00''$). Its extension
past in time on MDM data shows no variations. The good sampling
along the bump allows to get a rather robust
estimation of the Einstein time, $t_E=13\pm 4$ days.
This value, together with its position, makes
the bulge-bulge self-lensing hypothesis
the most likely for this event. 

PA-00-S4: This event is found far away from the M31 center,
but only at $2'\,54''$ from the center of the dwarf galaxy
M32. A detailed analysis\cite{point02} shows that the source
is likely to be a M31 disk A star, the main evidence being
the observed rather blue colour $(R-I) = 0.0\pm 0.1$. Given that
M32 lies $\sim$ 20 kpc in front of M31, the study of the relative
optical depth allows to conclude that the most likely
position for the lens is M32. 

\section{Conclusions} 

We have presented the results of microlensing survey
towards LMC by using the new picture 
of LMC given by van der Marel et al. \cite{marel01a}.
One interesting feature, that clearly emerges in this framework
by studying the microlensing signature we expect to find,
is an evident near--far asymmetry of the optical depth for 
lenses located in the LMC halo. 
Indeed, similarly to the case of M31 \cite{crotts92,jetzer94}, 
and as first pointed out by Gould\cite{gould93},
since the LMC disk is inclined, the optical depth is higher along 
lines of sight passing through larger portions of the LMC halo.
Such an asymmetry is not expected, on the contrary,
for a self-lensing population of events. 
What we show is that, indeed, a spatial asymmetry that goes
beyond the one expected from the observational strategy alone,
and that is coherent with that expected because of the inclination
of the LMC disk, is actually present. With the care suggested
by the small number of detected events on which this analysis
is based, this can be looked at, as yet observed by Gould\cite{gould93}, as a
signature of the presence of an extended halo around LMC.

As already remarked, any spatial asymmetry is \emph{not}
expected for a self-lensing population of events,
so that what emerges from this analysis can be considered
as an argument to exclude it. 
 
Furthermore, keeping in mind the observation\cite{evans}
that the timescale distribution of 
the events and their spatial variation across the
LMC disk offers possibilities of identifying the dominant lens
population, we have carefully characterized the ensemble 
of observed events under the hypothesis that
all of them do belong to the self-lensing population.
Through this analysis we have been able
to identify a large subset of events that
can not be accounted as part of this population.
Again, the small amount of events at disposal does not yet
allow us to draw sharp conclusions, although, the various arguments 
mentioned above are all consistent among them and converge quite clearly
in the direction of excluding self-lensing as being the major
cause for the events.

Once more observations will be available, as will hopefully
be the case with the SuperMacho experiment under way \cite{stubbs}, 
the use of the above outlined  methods can bring to a definitive answer to the problem
of the location of the MACHOs and thus also to their nature.

As a general outcome of presently available
pixel lensing results, we can clearly infer that
the detection of microlensing events towards M31 is now established.
The open issue to be still explored is the 
study of the M31 halo fraction in form of MACHOs.
With respect to this analysis, the events detected so far are all compatible with
stellar lenses, but the MACHO hypothesis is still open,
and we recall that the analysis for the INT data is still not concluded
(besides a third year data, variations with $\Delta R_{max}>21$ 
have still to be studied). Once this analysis completed, it is the necessary to ``weight''
it with an efficiency study of the pipeline of detection 
before meaningfully comparing its results with the prediction
of a Monte Carlo simulation. This should eventually
allow us to draw firm conclusions
on the halo content in form of MACHO of M31.

------------------------------------------------------------------

\end{document}